\documentclass[aps,prd,twocolumn,groupedaddress]{revtex4}

\usepackage{graphicx}
\usepackage{dcolumn}
\usepackage{array,multirow}
\usepackage{bm}
\usepackage{amsmath}
\usepackage{epstopdf}
\usepackage{amsfonts}
\usepackage{amssymb}
\usepackage{epsfig}
\usepackage{tabularx}
\usepackage{color}
\usepackage[colorlinks=true,citecolor=blue,urlcolor=magenta,breaklinks]{hyperref}

\newcommand{\be}{\begin{equation}}
\newcommand{\en}{\end{equation}}
\newcommand{\bea}{\begin{eqnarray}}
\newcommand{\ena}{\end{eqnarray}}

\begin{document}

\title{Dark stars in Starobinsky's model}

\author{Grigoris Panotopoulos and  Il\'idio Lopes}
\email[]{grigorios.panotopoulos@tecnico.ulisboa.pt}
\email[]{ilidio.lopes@tecnico.ulisboa.pt}
\affiliation{Centro de Astrof{\'i}sica e Gravita\c c\~ao - CENTRA, Departamento de F{\'i}sica, Instituto Superior T{\'e}cnico - IST, Universidade de Lisboa - UL,
Av. Rovisco Pais 1, 1049-001 Lisboa, Portugal
}

\date{\today}

\begin{abstract}
In the present work we study non-rotating dark stars in $f(R)$ modified theory of gravity. In particular, we have considered bosonic self-interacting dark matter modelled inside the star as a Bose-Einstein condensate, while as far as the modified theory of gravity is concerned we have assumed Starobinsky's model $R+aR^2$. We solve the generalized structure equations numerically, and we obtain the mass-to-ratio relation for several different values of the parameter $a$, and for two different dark matter equation-of-states. Our results show that the dark matter stars become more compact in the R-squared gravity compared to General Relativity, while at the same time the highest star mass is slightly increased in the modified gravitational theory. The numerical value of the highest star mass for each case has been reported.
\end{abstract}

\maketitle


\section{Introduction}

A plethora of current observational data coming from many different sides show that we live in a spatially flat Universe that expands in an accelerating rate dominated by a dark sector, while ordinary stuff we are familiar with,
such as photons, neutrinos and baryons, comprise only a 5\% of the energy density of the Universe \cite{turner}. The dark sector, consisting of dark matter and dark energy, is one of the biggest challenges of modern theoretical cosmology, as its origin and nature still remain a mystery.

The concordance cosmological model, based on cold (collisionless) dark matter and a cosmological constant, has been very successful as it is in excellent agreement with a vast amount of observational data. However, the cosmological constant problems have forced us to explore other possibilities, such as dynamical dark energy models \cite{dynamical} with a time varying equation-of-state parameter or modified gravity models, e.g. $f(R)$ theories of gravity \cite{modified1,modified2}, where the Ricci scalar $R$ in the Einstein-Hilbert term of General Relativity (GR) is replaced by a generic function. Furthermore, long time ago self-interacting dark matter was proposed in order to eliminate or at least alleviate some apparent conflicts between the collisionless dark matter paradigm and astrophysical observations \cite{self-interacting}.

% Ilidio
Although the main motivation nowadays to study $f(R)$ theories of gravity is to explain the current cosmic acceleration, the astrophysical implications of this class of theories should also be investigated. Normal main-sequence stars and compact stars have been used to test and constrain alternative gravitational theories~\cite{2012ApJ...745...15C,2011PhRvL.107c1101P},as well as dark matter models~\cite{LS2010,Kouvaris10,britoetal1,britoetal2,LS2012}.
Compact objects, such as neutron stars or strange quark stars, are the final fate of stars, and thanks to their extreme conditions that cannot be reached in  Earth-based experiments, comprise excellent cosmic laboratories to test and constrain alternative gravitational theories. First it was pointed out in \cite{nosolution} that $f(R)$ theories of gravity cannot support interior solutions of relativistic stars, and they are thus unacceptable. However, in \cite{langlois} the authors showed that whether or not relativistic stars exist or not depends on the behaviour of the trace of the matter energy-momentum tensor. Relativistic stars in $f(R)$ theories of gravity have also been studied in \cite{fin1,fin2,greg,kokkotas,kokkotas1,kokkotas2}.

Recently it was shown that compact stars made entirely of self-interacting dark matter may exist \cite{darkstars}. In that work bosonic dark matter with a short-range repulsive potential was modelled inside the star as a Bose-Einstein condensate. In this scenario a polytropic equation-of-state of the form $P(\epsilon)=K \epsilon^2$ was derived, where the constant $K$ was found to be \cite{darkstars}
\begin{equation}
K = \frac{2 \pi l}{m_\chi^3}
\end{equation}
with $m_\chi$ being the mass of the dark matter particle, while $l$ is the scattering length that determines the dark matter self-interaction cross section $\sigma_\chi = 4 \pi l^2$. The properties of compact stars made of ordinary matter admixed with condensed dark matter have been studied in \cite{boson1,boson2,boson3,boson4}, and similarly admixed with fermionic matter in \cite{fermion1,fermion2,fermion3,fermion4}. 

Boson stars, albeit in a different context, have been already studied in \cite{BS1,BS2,BS3,BS4,BS5,BS6} and more recently in \cite{BS7}. The maximum mass for bosons stars in non-interacting systems was found in \cite{BS1,BS2}, while in \cite{BS3,BS4} it was pointed out that self-interactions can cause significant changes. In \cite{BS5,BS6} the authors constrained the boson star parameter space using data from galaxy and galaxy cluster sizes.

It is the aim of this work to study condensed dark stars in $f(R)$ theories of gravity, and in particular in Starobinsky's model $f(R)=R+aR^2$ \cite{starobinsky}. From a theoretical point of view the R-squared gravity is well-motivated, since higher order in $R$ terms are natural in Lovelock theory \cite{lovelock}, and also higher order curvature corrections appear in the low-energy effective equations of Superstring Theory \cite{ramgoolam}. Our work is organized as follows: In the next section we present the model and the structure equations, while in section 3 we discuss our numerical results. Finally we finish concluding our work in the fourth section.

\section{Theoretical framework}

The model is defined by the following action in the so-called Jordan frame
\begin{equation}
S = \frac{1}{16 \pi G} \int d^4x \sqrt{-g} f(R) + S_M[\psi_i, g_{\mu \nu}]
\end{equation}
where $G$ is Newton's constant, $g_{\mu \nu}$ is the metric tensor, $R$ is the Ricci scalar, and $S_M$ is the matter
action that depends on the metric tensor and the matter fields $\psi_i$. To avoid pathological situations, such as ghosts and tachyonic instabilities, it is required that viable $f(R)$ theories satisfy the conditions \cite{modified1,modified2}
\begin{eqnarray}
\frac{d f}{d R} & \geq & 0 \\
\frac{d^2 f}{d R^2} & \geq & 0
\end{eqnarray}

In the following we shall be considering the Starobinsky's model $f(R)=R+aR^2$, where $a \geq 0$ is the only free parameter of the gravitational theory, and clearly the $a=0$ case corresponds to GR. The parameter $a$ has dimensions
$[mass]^{-2}$, and therefore it can be written also in the form $a=1/M^2$, where now the mass scale $M$ is the free parameter of the theory.

By performing a conformal transformation of the form \cite{woodard,davis,kokkotas,kokkotas1,kokkotas2}
\begin{equation}
\tilde{g}_{\mu \nu} = p g_{\mu \nu} = A^{-2} g_{\mu \nu}
\end{equation}
where $A(\phi)=exp(-\phi/\sqrt{3})$, the action can be equivalently written down in the more familiar Einstein frame
taking the form \cite{woodard,davis,kokkotas,kokkotas1,kokkotas2}
\begin{widetext}
\begin{equation}
S = \frac{1}{16 \pi G} \int d^4x \sqrt{-\tilde{g}} [\tilde{R}-2 \tilde{g}^{\mu \nu} \partial_\mu \phi \partial_\nu \phi-V(\phi)] + S_M[\psi_i, \tilde{g}_{\mu \nu} A(\phi)^2]
\end{equation}
\end{widetext}
where the system looks like GR with an extra scalar field with a self-interaction potential $V(\phi)$ given by \cite{woodard,kokkotas1,kokkotas2}
\begin{equation}
V(\phi) = \frac{(p-1)^2}{4 a p^2} = \frac{(1-exp(-2 \phi/\sqrt{3}))^2}{4 a}
\end{equation}
Varying with respect to the metric tensor and the scalar field one obtains Einstein's field equations as well as the Klein-Gordon equation
\begin{eqnarray}
\tilde{G}_{\mu \nu} & = & 8 \pi G [ \tilde{T}_{\mu \nu} + T^{\phi}_{\mu \nu} ] \\
\nabla_\mu \nabla^\mu{\phi}-\frac{1}{4} V_{,\phi} & = & -4 \pi G \alpha \tilde{T}
\end{eqnarray}
where $T^{\phi}_{\mu \nu}$ is the stress-energy tensor corresponding to the scalar field, $,\phi$ denotes differentiation with
respect to the scalar field, while due to the conformal transformation there is a direct coupling between matter and the scalar field with the coupling constant being $\alpha=-1/\sqrt{3}$ \cite{kokkotas,kokkotas1,kokkotas2}.
The matter energy-momentum tensor $T_{\mu \nu}$ in the Jordan frame and $\tilde{T}_{\mu \nu}$ in the Einstein frame are 
related via \cite{kokkotas}
\begin{equation}
\tilde{T}_{\mu \nu} = A(\phi)^2 T_{\mu \nu}
\end{equation}
and in particular in the case of a perfect fluid the energy densities and the pressures in the two frames are related 
via \cite{kokkotas}
\begin{eqnarray}
\tilde{P} & = & A(\phi)^4 P \\
\tilde{\epsilon} & = & A(\phi)^4 \epsilon
\end{eqnarray}
where the tilde indicates the Einstein frame. Finally, for static spherically symmetric solutions of the form
\begin{equation}
ds^2 = -e^{2 \nu(r)} dt^2 + e^{2 \lambda(r)} dr^2 + r^2 (d \theta^2 + sin^2 \theta d \varphi^2)
\end{equation}
one obtains the following structure equations \cite{kokkotas}
\begin{widetext}
\begin{eqnarray}
\frac{1}{r^2} \frac{d}{d r} [r (1-exp(-2 \lambda))] & = & 8 \pi G A(\phi)^4 \epsilon + \frac{V(\phi)}{2} + exp(-2 \lambda) \left( \frac{d \phi}{d r} \right)^2 \\
\frac{2}{r} exp(-2 \lambda) \frac{d \nu}{d r}-\frac{1}{r^2} (1-exp(-2 \lambda)) & = & 8 \pi G A(\phi)^4 P - \frac{V(\phi)}{2} + exp(-2 \lambda) \left( \frac{d \phi}{d r} \right)^2 \\
\frac{d^2 \phi}{dr^2} + \left( \frac{2}{r}+\frac{d \nu}{d r}-\frac{d \lambda}{d r} \right) \frac{d \phi}{d r} & = & 4 \pi G \alpha A(\phi)^4 (\epsilon-3 P) exp(2 \lambda) + \frac{1}{4} V_{,\phi} exp(2 \lambda) \\
\frac{d P}{d r} & = & - (P+\epsilon) \left( \frac{d \nu}{d r} + \alpha \frac{d \phi}{d r} \right)
\end{eqnarray}
\end{widetext}
Clearly when the scalar field is absent one recovers the usual
Tolman-Oppenheimer-Volkoff equations \cite{TOV}. The system of coupled equations is supplemented with the EoS $P(\epsilon)=K \epsilon^2$ as well as with the initial conditions at the center of the star
\begin{eqnarray}
\epsilon(0) & = & \epsilon_c \\
\lambda(0) & = & 0 \\
\phi(0) & = & \phi_c \\
\frac{d \phi}{d r}(0) & = & 0
\end{eqnarray}
where $\epsilon_c$ and $\phi_c$ are the central values of the dark matter energy density and of the scalar field respectively, and the last condition ensures the regularity of the scalar field. In principle one could handle the problem in perturbations theory, assuming that the parameter $a$ is small and therefore the higher order term in $R$ just perturbs the GR solution. However, it was shown in \cite{kokkotas} that in dealing with relativistic stars an analysis based on perturbation theory does not provide us with reliable results, and therefore in the present work we shall treat the problem exactly, i.e. non-perturbatively.
We wish to stress the fact that contrary to $\epsilon_c$, $\phi_c$ is not an arbitrary initial value but it must be determined self-consistently. In the next section we explain how this is done.

\section{Numerical results}

Before we discuss the numerics and present our findings let us first discuss the limits that current data put on the free parameters of the model, namely $M$ and $K$. First, it is known that a light scalar field can mediate a long range attractive force leading to a modification to the usual Newtonian potential $1/r$. Thus, to avoid solar system tests the scalar field must be heavy enough. In \cite{postnewtonian} the authors obtained the expression for the Post-Newtonian parameter $\gamma$ in $f(R)$ theories of gravity, which has been measured by the Cassini spacecraft. From the one hand the parameter $\gamma$ in $f(R)$ theories it is computed to be
\begin{equation}
\gamma = \frac{3-exp(-m r)}{3+exp(-m r)}
\end{equation}
where $m$ is the mass of the scalar field, which in Starobinsky's model is given by $m=M/\sqrt{6}$ \cite{kokkotas,postnewtonian}. On the other hand the Cassini mission provides us with the measurement \cite{cassini}
\begin{equation}
\gamma = 1 + (2.1 \pm 2.3) \times 10^{-5}
\end{equation}
for which $r=1.5 \times 10^8 km$. Therefore the mass scale $M$ characterizing Starobinsky's model must satisfy the bound
\begin{equation}
M \geq 4 \times 10^{-26} GeV
\end{equation}
In addition, the Starobinsky model is a viable candidate for cosmological inflation, and according to the latest Planck data $a \simeq 10^{-45} (N/50)^2 eV^{-2}$ \cite{staro,gannouji}, where $N$ is the number of e-folds.

\begin{figure}[ht!]
\centering
\includegraphics[scale=0.7]{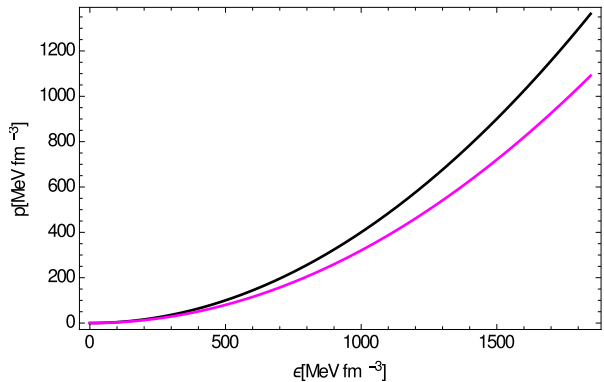}
\caption{Pressure versus energy density for two dark matter models.}
	\label{fig:1} 	
\end{figure}

Furthermore, self-interacting dark matter is constrained by current observations which require that \cite{bullet1,bullet2,review}
\begin{equation}
0.45 \frac{cm^2}{g} < \frac{\sigma_\chi}{m_\chi} < 450 \frac{cm^2}{g}
\end{equation}
For a scattering length $l \sim 10 fm$ and for a dark matter mass $m_\chi \sim GeV$ it is possible to satisfy the above limits and at the same time obtain a value for $K \approx 50 GeV^{-4}$. Therefore in the following we
shall consider a) the following three different values of the $a$ parameter
\begin{eqnarray}
a_1 & = & \frac{5 \times 10^{76}}{m_{pl}^2} \\
a_2 & = & \frac{10^{77}}{m_{pl}^2} \\
a_3 & = & \frac{10^{78}}{m_{pl}^2}
\end{eqnarray}
and b) two dark matter EoSs, one stiff and one soft, shown in Fig.~\ref{fig:1}.

We remark in passing that these values of the $a$ parameter are also fine with
the emission of gravitational radiation from observed binary systems. For the calculation in the framework of GR see e.g. \cite{GR1,GR2}, while for the relevant calculation in $f(R)$ gravity see \cite{Laurentis1,Laurentis2,Laurentis3}. 

Regarding the DM equation-of-state the numerical values we have used are as follows. For the stiff EoS 
\begin{equation}
\textrm{Stiff EoS} \rightarrow
\left\{
\begin{array}{lcl}
m_\chi = 1.93 GeV \\
&
&
\\
l = 11.78 fm
\end{array}
\right.
\end{equation}
corresponding to the black curve in Fig.~\ref{fig:1}, while for the soft EoS
\begin{equation}
\textrm{Soft EoS} \rightarrow
\left\{
\begin{array}{lcl}
m_\chi = 2.11 GeV \\
&
&
\\
l = 12.32 fm
\end{array}
\right.
\end{equation}
corresponding to the magenta curve in Fig.~\ref{fig:1}.

A final remark is in order here regarding the solution of the exterior problem. It is known that $f(R)$ theories of gravity admit static spherically symmetric black hole solutions of the Schwarzschild-de Sitter form \cite{BHsol}, where the lapse function is given by $f(r)=1+c_1/r-R_0/12$, with $c_1$ being an arbitrary integration constant related to the mass of the black hole, and $R_0$ being the root of the algebraic equation $R f'(R)=2 f(R)$, while the prime denotes differentiation with respect to $R$. For Starobinsky's model $R_0=0$, and the black hole solution is precisely Schwarzschild without a cosmological constant. This can be easily seen using the equations presented before valid in the Einstein frame (without matter $P=0=\epsilon$) as follows. If we assume solutions with a constant value for the scalar field, $\phi=const=\phi_0$, that in addition corresponds to an extremum of the potential, $V_{,\phi}(\phi_0)=0$, then the two last equations are automatically satisfied, while the first two are essentially the tt and rr equations of GR with a cosmological constant $\Lambda=V(\phi_0)/2$ \cite{prototype}.
For the Starobinsky's model $\phi_0=0=V(\phi_0)$, and we thus obtain the anticipated Schwarzschild black hole solution. This implies that in the exterior problem solution all three quantities $P=\epsilon=\phi=0$, and therefore when we consider the interior problem solution we must require that both the dark matter energy density and the scalar field vanish on the surface of the star
\begin{eqnarray}
\epsilon(R_*) & = & 0 \\
\phi(R_*) & = & 0
\end{eqnarray}
which clearly is not possible for any $\phi_c$. Therefore for a given equation of state (or $K$ value), a given $f(R)$ model (or $a$ value) and a given $\epsilon_c$ the scalar field central value $\phi_c$ is determined requiring that $\phi,\epsilon$ vanish simultaneously at the surface of the star. Finally, as usual the condition $P(R_*)=0$ determines the radius of the star $R_*$, while the star mass $M_*$ is given by $M_*=m(R)$, where we have introduced a new function defined by
\begin{equation}
1-\frac{2 m(r)}{r}=e^{-2 \lambda(r)}
\end{equation}

\begin{figure}[ht!]
\centering
\includegraphics[scale=0.7]{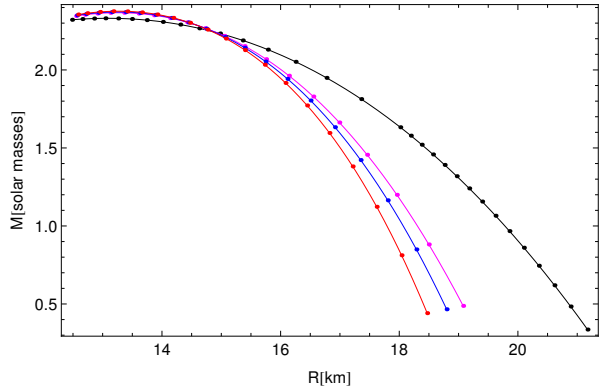}
\caption{Mass-to-radius relation for the stiff EoS (larger $K$) in GR (black) and for 3 different values of the parameter $a$.}
	\label{fig:2} 	
\end{figure}

\begin{figure}[ht!]
\centering
\includegraphics[scale=0.7]{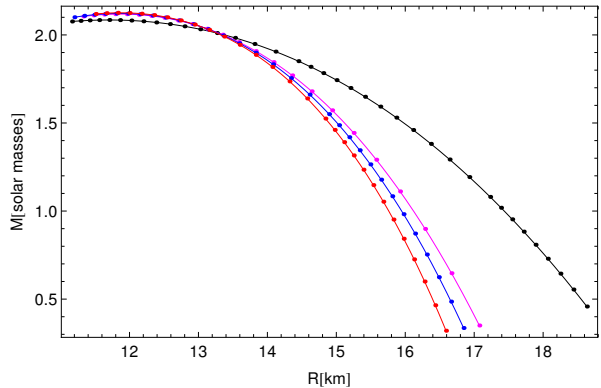}
\caption{Same as before, but for the soft EoS (lower $K$).}
	\label{fig:3} 	
\end{figure}

In Fig.~\ref{fig:2} we show the star mass (in solar masses) versus the star radius (in km) for the stiff EoS (black colour in Fig.~\ref{fig:1}) and for the 3 different values of $a$. For comparison,
the mass-to-ratio relation that corresponds to GR is also shown in black.
As we increase $a$ the curves are shifted downwards. For low radii the curves lie one very close to another and begin to separate at $R_* ~ 14.5 km$. For a given mass the DM star in GR is characterized by a larger radius, and therefore the higher order term in $R$ makes the star more compact. Fig.~\ref{fig:3} corresponds to the soft equation-of-state (magenta colour in Fig.~\ref{fig:1}). The pattern observed in the previous plot is repeated here. The part of the curves that corresponds to large radius and low mass is obtained for low central energy density, while the part of the curves that corresponds to high mass and low radius is obtained for high central energy density. The curves exhibit a maximum precisely in that part, and this corresponds to the highest star mass which increases with $a$. In Tables~\ref{table:First} and \ref{table:Second} we report the numerical values of the highest mass supported by each EoS in a given gravitational theory.

Note that since the second EoS is softer the highest star mass is lower, as it could have been anticipated. Also note that the causality requirement $c_s \leq 1$, with $c_s$ being the speed of sound, forces us to stop at a certain point, since for sufficiently high central energy density the speed of sound defined as
\begin{equation}
c_s^2 = \frac{dP}{d \epsilon}=2 K \epsilon
\end{equation}
exceeds unity. It is easy to verify that for a given constant $K$ the maximum allowed $\epsilon_c$ is given by $\epsilon_c^{max}=1/(2 K)$.

\begin{table}
\begin{tabular}{l | l}
Model & \multicolumn{1}{c}{{\sc Maximum $M_*$}}\\
      & $M_{\odot}$ \\
\hline
\hline
GR  &  2.33  \\
Model 1 & 2.37 \\
Model 2 & 2.37 \\
Model 3 & 2.38
\end{tabular}
\caption{Highest star mass (in solar masses) for the stiff EoS in GR and in Starobinsky's model for 3 different value of $a$.}
\centering
\label{table:First}
\end{table}

\begin{table}
\begin{tabular}{l | l}
Model & \multicolumn{1}{c}{{\sc Maximum $M_*$}}\\
      & $M_{\odot}$ \\
\hline
\hline
GR  &  2.08  \\
Model 1 & 2.12 \\
Model 2 & 2.12 \\
Model 3 & 2.13
\end{tabular}
\caption{Highest star mass (in solar masses) for the soft EoS in GR and in Starobinsky's model for 3 different value of $a$.}
\centering
\label{table:Second}
\end{table}

\section{Conclusions}

We have studied dark stars in Starobinsky's model $f(R)=R+aR^2$, and we have solved the structure equations numerically to produce the mass-to-radius relation for the relativistic stars. The dark matter is assumed to be bosonic and self-interacting modelled inside the star as a Bose-Einstein condensate leading to a polytropic EoS. We have worked in the Einstein frame where the presence of a scalar field modifies the standard TOV equations, and we have treated the higher order in $R$ term non-perturbatively. We have produced the $M-R$ diagram for two EoSs, one stiff and one soft, and for 3 different values of the parameter $a$ compatible with solar system tests. Our results show that the modified theory of gravity makes the relativistic stars even more compact, while at the same increase slightly the highest star mass, the values of which are reported.

%%%%%%%%%%%%%%%%%%%%%%%%%%%%%%%%%%%%%%%%%%%%%%%%%%%%%%%%%%%%%%%%%%%%%%%%%%%%%%%%%%%%%%%%%%%%%%%%%%%%%%%%%%%%%%%%%%%%%

\section*{Acknowlegements}

We wish to thank R. Gannouji for communications, and the anonymous reviewer for useful suggestions. The authors thank the Funda\c c\~ao para a Ci\^encia e Tecnologia (FCT), Portugal, for the financial support to the Center for Astrophysics and Gravitation-CENTRA,  Instituto Superior T\'ecnico,  Universidade de Lisboa,  through the Grant No. UID/FIS/00099/2013.

%%%%%%%%%%%%%%%%%%%%%%%%%%%%%%%%%%%%%%%%%%%%%%%%%%%%%%%%%%%%%%%%%%%%%%%%%%%%%%%%%%%%%%%%%%%%%%%%%%%%%%%%%%%%%%%%%%%%%%

\end{document}